\documentclass[fleqn,usenatbib]{mnras}

\usepackage{ulem}

\usepackage[T1]{fontenc}

\DeclareRobustCommand{\VAN}[3]{#2}
\let\VANthebibliography\thebibliography 

\def\thebibliography{\DeclareRobustCommand{\VAN}[3]{##3}\VANthebibliography}

\usepackage{graphicx}	
\usepackage{amsmath}	
\usepackage{amssymb}	
\usepackage{tabularx}

\title[Astrometric star-cluster membership probability]{
Astrometric star-cluster membership probability: 
application to the case of M\,37 with Gaia\,EDR3
}

\author[M.\ Griggio \& L.\ R.\ Bedin]{
M. Griggio$^{1,2}$\thanks{E-mail: massimo.griggio@inaf.it} and L. R. Bedin,$^{2}$
\\
$^{1}$Dipartimento di Fisica, Universit\`a di Ferrara, Via Giuseppe Saragat 1, Ferrara I-44122, Italy\\
$^{2}$Istituto Nazionale di Astrofisica, Osservatorio Astronomico di Padova, Vicolo dell'Osservatorio 5, Padova I-35122, Italy\\
}

\date{Accepted 2022 February 9. Received 2022 February 7; in original form 2021 December 28}

\pubyear{2022}

\usepackage{newtxtext,newtxmath}

\begin{document}
\label{firstpage}
\pagerange{\pageref{firstpage}--\pageref{lastpage}}
\maketitle

\begin{abstract}
In this work, starting from the well-accepted relations in literature, we introduce a new formalism to compute the astrometric membership probabilities for sources in star clusters, and we provide an application to the case of the open cluster M\,37. 
The novelty of our approach is a refined --and magnitude-dependent-- modelling of the parallax distribution of the field stars. 
We employ the here-derived list of members to estimate the 
cluster's mean systemic astrometric parameters, which are based on the 
most recent Gaia's catalog (EDR3). 
\end{abstract}





\section{Introduction}

Star clusters represent one of our most important sources of knowledge of stellar formation and evolution: the measurements of their distance, age and chemical composition provide strong constraints on astrophysical models of stellar evolution. Star clusters consist of gravitationally bound stars which share the same distance and center of mass motion, and they appear as a stellar over-density in a region of the sky. In the studies of these objects one of the most crucial steps is the determination of the membership probability of the observed stars, to distinguish actual members of the cluster from field stars that lie in the same region but are not bound to the cluster.

Traditionally, the problem of estimating membership probabilities using the astrometric parameters of the sources has been treated with techniques that were developed in the pioneering work by \cite{1958AJ.....63..387V} and \cite{1971A&A....14..226S}. In their works the distribution of sources in the vector-point diagram (VPD) is modelled as a mixture of two Gaussian distributions, one for the cluster members and another one for the field sources. This method was further refined by the contribution of several authors \citep[see][and refs therein]{1998ycat..41310089T,1998A&AS..133..387B}.

An additional improvement of this technique introduced by \cite{1995AJ....109..672K} foresees the partition of the data in brightness (a \textit{sliding window} in magnitude) and spatial bins when deriving the parameters of the distributions. One of the advantages of using a ``local sample'' approach is that membership probabilities are not biased by possible differences in the shape of the field and cluster luminosity functions, or in proper motion accuracy for bright and faint stars.

In this work we discuss an improvement of the astrometric method exploiting the Gaia astrometry to increase the separation between cluster and field stars.

Including parallaxes provides additional information to estimate membership probabilities. 
While multiple publications since 1998 
have taken into account Hipparcos \citep[e.g.][]{1999A&A...345..471R,2000A&AS..146..251B}, 
and later Gaia \citep[e.g.][]{2018ApJ...856...23G,2018A&A...615A..49C,2018A&A...618A..59C,2020MNRAS.499.1874M} parallaxes, none of these works introduced a proper formalism, with the only exception of \cite{2020MNRAS.499.1874M}, which however made an oversimplification that will be discussed later.
\\

This paper is organized as follows: in Section \ref{sec1} we review the classical formalism used to compute the membership probability, in Section \ref{sec2} we introduce the new term to account for the parallax distribution, while in Section \ref{sec3} we compare the membership calculated with this new term and without it, taking the open cluster M\,37 as a test case.
In Sections \ref{sec4} and \ref{sec5} 
we use the membership probability to select a list of cluster's members and we use them to derive a new estimate of the cluster's mean proper motion and parallax. We also publicly release a catalog of all the sources with the membership probabilities.
Finally, in Section \ref{sec6} we provide a summary of this work.


\section{Membership probability: the classical approach}
\label{sec1}

In this section we will review the formalism ``traditionally'' employed to determine the membership probability of the $i$-th star using four out of its five astrometric parameters, namely its position $(x_i,y_i)$ and its proper motion $(\mu_{x_i},\mu_{y_i})$. We will follow the formulation from \cite{1998ycat..41310089T} and \cite{1998A&AS..133..387B}.

In these works the cluster membership probability of the $i$-th star is calculated as
\begin{equation}
    P_{\rm c}(i)=\frac{\Phi_{\rm c}(i)}{\Phi(i)},
\end{equation}
where $\Phi_{\rm c}$ is the cluster distribution function and $\Phi$ is the total distribution given by 
\begin{equation}
\Phi=\Phi_{\rm c}+\Phi_{\rm f},
\end{equation}
with $\Phi_{\rm f}$ the distribution function of field stars. The distribution function of cluster (and field) stars is given by the contribution of two terms, i.e.,  
\begin{equation}
    \label{phi}
    \Phi_{\rm c/f} = n_{\rm c/f} \cdot \Phi_{\rm c/f}^\upsilon \cdot \Phi_{\rm c/f}^r,
\end{equation}
in which $\Phi^\upsilon$ is the distribution function in the velocity space, $\Phi^r$ is the distribution function in the position space and $n$ is the normalized number of stars ($n_{\rm c} + n_{\rm f}=1$). 

For the cluster velocity distribution they adopt an asymmetric 2D Gaussian in the form:
\begin{align}
    \label{phicv}
    \Phi_{\rm c}^\upsilon (i) = &\frac{1}{2\pi ( \sigma_{\mu_{x_{ \rm c}}}^2+\epsilon_{\mu_{x_{i}}}^2)^{1/2}( \sigma_{\mu_{y_{\rm c}}}^2+\epsilon_{\mu_{y_{i}}}^2 )^{1/2} } \nonumber \\
    &\exp \left\{ -\frac{1}{2} \left[ 
    \frac{\left(\mu_{x_{i}}-\mu_{x_{\rm c}}\right)^2}{\sigma_{\mu_{x_{ \rm c}}}^2+\epsilon_{\mu_{x_i}}^2} +  
    \frac{\left(\mu_{y_{i}}-\mu_{y_{\rm c}}\right)^2}{\sigma_{\mu_{y_{ \rm c}}}^2+\epsilon_{\mu_{y_i}}^2} \right] \right\} ,
\end{align}
where $\left( \mu_{x_i},\mu_{y_i}\right)$ are the proper motions of the $i$-th star, $\left( \mu_{x_{ \rm c}},\mu_{y_{\rm c}}\right)$ is the cluster proper motion center,
$( \sigma_{\mu_{x_{ \rm c}}},\sigma_{\mu_{y_{ \rm c}}} )$
is the intrinsic proper motion dispersion of member stars and 
$( \epsilon_{\mu_{x_i}},\epsilon_{\mu_{y_i}} )$ 
are the observed errors of the proper motions of the $i$-th star. Similarly, for the field stars velocity distribution we have
\begin{align}
    \label{phifv}
    \Phi_{\rm f}^\upsilon (i) = &\frac{ 1 }
    {2\pi \left(1-\gamma^2\right)^{1/2} (\sigma^2_{\mu_{x_{ \rm f}}}+\epsilon^2_{\mu_{x_i}} )^{1/2} (\sigma^2_{\mu_{y_{ \rm f}}}+\epsilon^2_{\mu_{y_i}} )^{1/2} } \nonumber \\
    &\exp \Bigg\{  -\frac{1}{2\left(1-\gamma^2\right)} \Bigg[ \frac{\left(\mu_{x_i}-\mu_{x_{ \rm f}}\right)^2}{\sigma_{\mu_{x_{ \rm f}}}^2+\epsilon_{\mu_{x_i}}^2} \nonumber \\
     &- \frac{2\gamma\left(\mu_{x_i}-\mu_{x_{ \rm f}}\right)\left(\mu_{y_i}-\mu_{y_{ \rm f}}\right) }{(\sigma_{\mu_{x_{ \rm f}}}^2+\epsilon_{\mu_{x_i}}^2)^{1/2}(\sigma_{\mu_{y_{ \rm f}}}^2+\epsilon_{\mu_{y_i}}^2)^{1/2}}
    + \frac{\left(\mu_{y_i}-\mu_{y_{ \rm f}}\right)^2}{\sigma_{\mu_{y_{ \rm f}}}^2+\epsilon_{\mu_{y_i}}^2}
    \Bigg]  \Bigg\},
\end{align}
where $\left( \mu_{x_i},\mu_{y_i}\right)$ are the proper motions of the $i$-th star, 
$\gamma$ is the correlation coefficient between $\mu_{x_i}$ and $\mu_{y_i}$, 
$\left( \mu_{x_{ \rm f}},\mu_{y_{ \rm f}}\right)$ the field proper motion center, 
$( \epsilon_{\mu_{x_i}},\epsilon_{\mu_{y_i}})$ the observed errors of the proper motions of the $i$-th star and 
$( \sigma_{\mu_{x_{ \rm f}}},\sigma_{\mu_{y_{ \rm f}}} )$ the field intrinsic proper motion dispersion.

For the spatial distribution of cluster members a simple (and sufficient for the purpose) approximation is to  use a Gaussian profile:
\begin{equation}
    \label{phir}
    \Phi_{\rm c}^r (i) = \frac{1}{2\pi r_{\rm c}^2}
    \exp \left\{ -\frac{1}{2} \left[ \left( \frac{x_i - x_{\rm c}}{r_{\rm c}} \right)^2 
    + \left( \frac{y_i - y_{\rm c}}{r_{\rm c}} \right)^2\right]  \right\},
\end{equation}
in which $\left( x_i,y_i\right)$ is the position of the $i$-th star, $\left( x_{\rm c},y_{\rm c}\right)$ the center of the cluster and $r_{\rm c}$ the characteristic radius. The field star spatial distribution is assumed to be flat:
\begin{equation}
    \label{phirf}
    \Phi_{\rm f}^r (i) = \frac{1}{\pi r^2_{\rm max}},
\end{equation}
where $r_{\rm max}$ is the radius of the portion of the sky under exam (assuming it has a circular shape).

This method to compute the membership probabilities was applied in a number of papers in the recent literature \citep[see for example][]{2008A&A...484..609Y,2009A&A...493..959B,2018MNRAS.481.3382N,2021MNRAS.505.3549S}.

\section{Including the parallax}
\label{sec2}

The Gaia\,EDR3 \citep{2016A&A...595A...1G,2021A&A...649A...1G} catalog is an unprecedented astronomical data set in terms of its size and astrometric precision and accuracy. In particular, it provides the full 5-parameter astrometric solution (positions, proper motions and parallaxes) and magnitudes in its three photometric bands ($G$, $G_{BP}$, $G_{RP}$) for more than 1.4 billion sources, with a limiting magnitude of about $G \approx 21$ and a bright limit of about $G \approx 3$. Thanks to the Gaia\,EDR3 exquisite astrometry we can extend the formalism presented in the previous section including a new term to take into account the parallax distribution. Particularly, the parallax uncertainties in the EDR3 are $0.02-0.03$\,mas for $G<15$, $0.07$\,mas at $G=17$, $0.5$\,mas at $G=20$ and $1.3$\,mas at $G=21$ \citep{2021A&A...649A...2L}. This unmatched level of precision allows us to include the parallax in the computation of the membership probability, thus achieving a more robust estimate for this fundamental quantity. 

To account for the parallax distribution we rewrite Equation \ref{phi} as:
\begin{equation}
\label{phipi}
    \Phi_{\rm c/f} = n_{\rm c/f} \cdot \Phi_{\rm c/f}^\upsilon \cdot \Phi_{\rm c/f}^r \cdot \Phi_{\rm c/f}^\varpi,
\end{equation}
where $\Phi_{\rm c/f}^\varpi$ is the distribution function of the parallaxes for the cluster members and for the field stars.

We can assume that the parallaxes of cluster members are normally distributed, such that: 
\begin{equation}
    \label{phipc}
    \Phi_{\rm c}^\varpi (i) = \frac{1}{ \left(2\pi ( \sigma_{\varpi_{\rm c}}^2 + \epsilon_{\varpi_{i}}^2  )\right)^{1/2} }
    \exp \left\{ -\frac{1}{2} \left( \frac{\varpi_i - \varpi_{\rm c}}
    {\sigma_{\varpi_{\rm c}}^2 + \epsilon_{\varpi_i}^2} \right)^2  \right\},
\end{equation}
where $\left(\varpi_i,\varpi_{\rm c}\right)$ are the parallax of the $i$-th star and of the cluster respectively, $\epsilon_{\varpi_i}$ are the observed errors of the parallax of the $i$-th star and $ \sigma_{\varpi_{\rm c}}$ the cluster intrinsic parallax dispersion (in the case where the size of the cluster is not negligible compared to its distance).

However, modeling the distribution function of the parallaxes of field stars, $\Phi_{\rm f}^\varpi$, is more complicated: we are not observing an ensemble of stars all at the same distance, or at an average distance with a normal distribution around the mean. In the case of the parallaxes of the field we are rather observing the closest stars to the Sun and stars potentially well into the Galactic Halo. The exact distribution function of stars in the Galactic field in different directions and at the various magnitudes is hard to model, and to derive an accurate distribution is well beyond the purpose of this paper. 
For our purposes, it will be sufficient to adopt a simple approximation, analogous to what described in Eq.\,\ref{phifv} for the proper motion distribution of field objects $\Phi_{\rm f}^\upsilon$ (which is a widely accepted approximation in literature).
To reproduce the field stars' parallax distribution we adopted a sum of two Gaussian functions, which are assumed to model as well the measurement errors in the parallaxes. This choice let us reproduce very well the parallax at each magnitude bin without complicating too much our formalism. Therefore, we have: 
\begin{align}
    \label{phipf}
    \Phi_{\rm f}^\varpi = &\frac{A_1}{ (2\pi  \sigma_{1_{\rm f}}^2)^{1/2} }
    \exp \left\{ -\frac{1}{2} \left( \frac{\varpi_i - \varpi_{1_{\rm f}}}
    {\sigma_{1_{\rm f}}^2} \right)^2  \right\} \nonumber \\
    &+ \frac{A_2}{ (2\pi \sigma_{2_{\rm f}}^2 )^{1/2} }
    \exp \left\{ -\frac{1}{2} \left( \frac{\varpi_i - \varpi_{2_{\rm f}}}
    {\sigma_{2_{\rm f}}^2 } \right)^2  \right\},
\end{align}
where $A_1$, $A_2$, $\varpi_{1_{\rm f}}$, $\varpi_{2_{\rm f}}$, $\sigma_{1_{\rm f}}$ and $\sigma_{2_{\rm f}}$ are the parameters of the two Gaussian and $\varpi_i$ the parallax of the $i$-th star.
We verified that this simple model for the distribution of the parallaxes for field objects is a general valid approximation.  
To this aim, we downloaded portions of the Gaia\,EDR3 catalog in various directions of the sky, to probe different parts of the Galactic field. 
The obtained Gaia\,EDR3 distributions of the parallaxes for field objects at various magnitudes were always represented -- within the statistical sampling errors -- by our simple model.

We note here that \cite{2020MNRAS.499.1874M} followed a method that is qualitatively similar to ours, but they do not use the sliding window approach and they adopted a single Gaussian model; while this assumption works well in the case of parallaxes dominated by errors (faint stars), it does not represents well the intrinsic parallax distribution of field stars when uncertainties are small.
We show a detailed comparison of the two models in the next section. 
%


%
\section{Example: the case of M 37}
\label{sec3}
\defcitealias{2020A&A...633A..99C}{CG20}

We considered the open cluster M\,37 (NGC\,2099) as a test case for this new formalism for the computation of the membership probabilities. We downloaded a portion of the Gaia\,EDR3 catalog centered on 
M\,37\,\citep[$\alpha_{\rm c}=88.074$\,deg, $\delta_{\rm c}=+32.545$\,deg,][hereafter CG20]{2020A&A...633A..99C} 
with a radius of $1.5$\,deg, and we computed the membership probability for each source both including and neglecting the contribution from the Gaia parallaxes. We adopted a sliding window in magnitude of $1.5$\,mag, which we found as a good compromise between having a good statistics at all magnitudes and considering sources with magnitude similar to the target star.
As initial guess, we employed for the cluster's systemic proper motion\footnote{Where for conciseness in the notation we indicate $\mu_{\alpha\cos{\delta}_{\rm c}}$ with $\mu_{\alpha_c}$ } 
$(\mu_{x_{ \rm c}},\mu_{y_{ \rm c}})=(\mu_{\alpha_c},\,\mu_{\delta_c})=(1.924,\,-5.648,)$\,mas/yr and for the systemic parallax $\varpi_{\rm c}=0.666$\,mas, which are the values given by \citetalias{2020A&A...633A..99C}.

We started by estimating the intrinsic dispersion of the proper motions of the cluster, i.e. $\sigma_{\mu_{x_{ \rm c}}}$ and $\sigma_{\mu_{y_{\rm c}}}$ of Eq.\,\ref{phicv}. We selected the members of M\,37 for which \citetalias{2020A&A...633A..99C} give their clustering score equal to one (the highest score), choosing only the sources with $G<17$ (where the Gaia errors are of the order, $10^{-2}$\,mas/yr). We $\sigma$-clipped the values at 3-$\sigma$ around the median, and then we calculated the $68.27^{\rm{th}}$ percentile of the residuals from the median of $\mu_{\alpha}$ and $\mu_{\delta}$, which we assumed as the observed dispersion. Subtracting in quadrature from the observed dispersion the median observational relative errors provided by Gaia\,EDR3 gives a reasonable estimate of the cluster intrinsic dispersion. We obtain $\sigma_{\mu_{x_{ \rm c}}}=\sigma_{\mu_{\alpha}}=137$\,$\mu$as and $\sigma_{\mu_{y_{ \rm c}}}=\sigma_{\mu_{\delta}}=138$\,$\mu$as.
Note that at a corresponding distance of 1.5\,kpc (for $\varpi=0.666$\,mas), these translate into transverse velocities of less than 1\,km/s, which is a reasonable value for such an open cluster \citepalias[e.g. see][and refs therein]{2020A&A...633A..99C}.
The values for proper motions and estimated errors of individual sources in the Eq.\,\ref{phicv} are taken straight from the Gaia\,EDR3 catalog, i.e.: 
$\mu_{x_i}=\mu_{\alpha_{i}}$, 
$\mu_{y_i}=\mu_{\delta_{i}}$, 
$\epsilon_{x_i}=\epsilon_{\mu_{\alpha_i}}$, and
$\epsilon_{y_i}=\epsilon_{\mu_{\delta_i}}$.  

The parameters $\mu_{x(y)_{\rm f}}$ and $\sigma_{\mu_{x(y)_{\rm f}}}$ of Eq.\,\ref{phifv} have been estimated from the sources in the magnitude window of the star under exam. We adopted as $\mu_{x(y) \rm f}$ the median values of the proper motion of field objects, and the $68.27^{\rm{th}}$ percentile of the residuals around these median as the observed dispersion: 
$\sigma^{\rm obs}_{\mu_{{x(y)_{ \rm f}}}}$. 
We then calculated the average observational error of the sources in the magnitude window, 
$\epsilon_{\mu_{x(y)_{\rm f}}}$, 
by clipping the errors at 3-$\sigma$ and computing the median. This value is then used to calculate the intrinsic dispersion for field proper motion as: $\sigma^2_{\mu_{x(y)_{\rm f}}} =  (\sigma^{\rm obs}_{\mu_{\rm _{x(y)_{\rm f}}}})^2 - \epsilon^2_{\mu_{x(y)_{ \rm f}}}$, which are the ones to be used in Eq.\,\ref{phifv}. Again, for individual sources the values for proper motions and proper motion errors in the Eq.\,\ref{phifv} are taken from the Gaia\,EDR3 catalog.

To deal with the spatial distribution, 
we projected the Gaia coordinates ($\alpha_i,\delta_i)$ on the tangent plane ($\xi,\eta$), adopting the center of the cluster as tangent point ($\alpha_{\rm c},\delta_{\rm c}$), employing standard relations \citep[e.g., see Eq.\,3 in ][]{2018MNRAS.481.5339B}. 
Therefore, the coordinates on the tangent plane 
became $x_i=\xi_i$ and $y_i=\eta_i$ in Eq.\,\ref{phir}. 
The estimate of $r_c$ of Eq.\, \ref{phir} have been performed from the stars in the magnitude bin of the target; we calculate $r_x$ ($r_y$) as the $68.27^{\rm th}$ percentile of the residuals from $x_c$ ($y_c$), and we adopt as cluster radius in the magnitude bin $r_c^2=r_x^2+r_y^2$. This procedure allows us to account for the different distributions of the stars in each magnitude bin, which is a proxy for different mass-bins (at least along the Main Sequence).
The parameter $r_{\rm max}$ of Eq.\,\ref{phirf} is the radius of the Gaia EDR3 slice that we considered, i.e. $r_{\rm max}=1.5$\,deg.

The intrinsic dispersion of the parallaxes ($\sigma_{\varpi_{\rm c}}$ in Eq.\,\ref{phipc}) is negligible for M\,37 and it can be set equal to zero. Again, for this initial computation we adopt for the cluster average parallax the value from \citetalias{2020A&A...633A..99C}, while the values of the parallax ($\varpi_i$) and parallax error ($\epsilon_{\varpi_{i}}$) for individual sources are those given by the Gaia\,EDR3 catalog.
In the next sections we will then derive our own estimates for the cluster mean proper motions and parallax, employing Gaia\,EDR3 instead of the DR2, and finally re-compute the updated membership probabilities.

\begin{figure*}
    \centering
    \includegraphics[width=.9\textwidth]{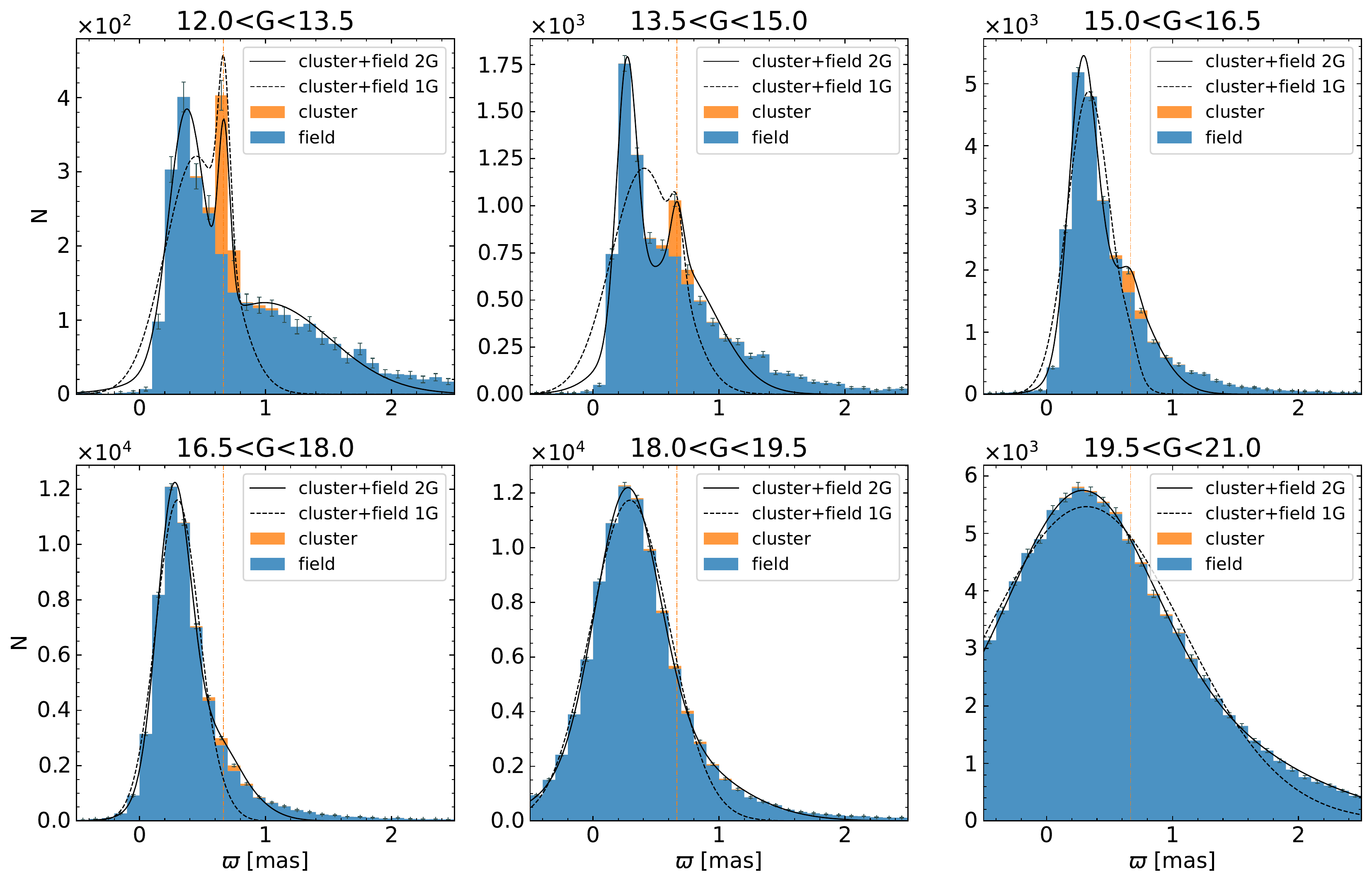}
    \caption{Parallax distribution in different magnitude bins in the M\,37 field of view. We plotted in orange the stars with parallax within 0.2\,mas from the cluster's mean parallax and proper motions within a circle of 0.5\,mas/yr in radius centered on the cluster's mean proper motions. Mean values were obtained from \protect\cite{2020A&A...633A..99C}. The solid black line represents the sum of the Gaussian of the cluster and the field modelled with two Gaussian functions, while the dashed black line is the same but using a single Gaussian to model the field's parallax distribution.}
    \label{fig:hist}
\end{figure*}

For the distribution of parallaxes in the Galactic field we used Eq.\,\ref{phipf}, so we fitted the distribution in each magnitude window with a sum of two Gaussian functions. From the fit we obtain the $\tilde{\sigma}_{1(2)_{\rm f}}$ parameters, which contain also the contributions of errors in parallaxes at the considered magnitude. To account for the observational errors, we compute the quantity $\sigma'^2_{1(2)_{\rm f}} = \tilde{\sigma}^2_{1(2)_{\rm f}} - \bar{\epsilon}^2_{1(2)_{\rm f}}$, where $\bar{\epsilon}_{1(2)_{\rm f}}$ is the median error of the stars in the magnitude window that we are considering, calculated after performing a 3-$\sigma$ clip. These are the values employed in Eq.\,\ref{phipf}, which are the sum in quadrature of intrinsic distributions and errors, $\sigma^2_{1(2)_{\rm f}} = \sigma'^2_{1(2)_{\rm f}} + \epsilon^2_{\varpi_{i}}$. 
We then used the Gaia\,EDR3 values for the parallax $\varpi_i$ and parallax error $\epsilon_{\varpi_{i}}$.
In Figure \ref{fig:hist} we show the distribution of the parallaxes in the M\,37 field of view, for different magnitude bins, where for comparison we show the fitted distribution of 
the parallaxes of cluster$+$field stars obtained in the case of a two-Gaussian model (solid line) and in the case of one-Gaussian model (dotted line).\\~ 

We finally computed the membership probability, hereafter $P_\varpi$, following Eq.\,\ref{phipi}, which includes the parallax, and we compared it with the membership probability calculated without including the parallax, i.e. $P$, computed according to Eq.\,\ref{phi}. 
The results show that including the parallax term allow us to better separate the cluster members from the field stars, in particular at fainter magnitudes ($16<G<18$).

\begin{figure}
    \centering
    \includegraphics[width=\columnwidth]{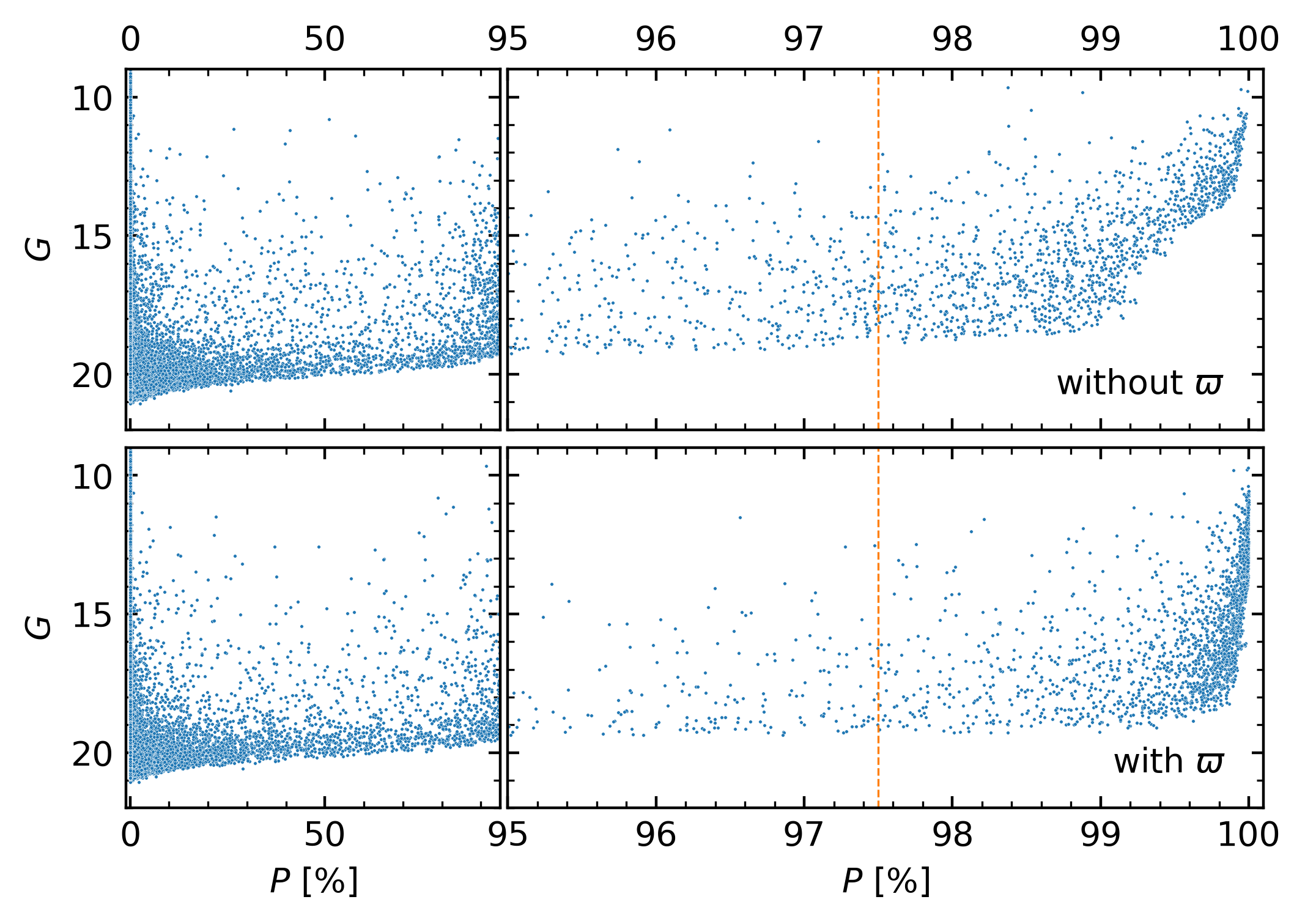}
    \caption{Comparison between the membership probability calculated without the parallax term (top) and accounting for the parallax of the sources (bottom); $P>95$\,\% are zoomed on the right panels. Vertical orange line indicates $P=97.5$\,\%.}
    \label{fig:mpg}
\end{figure}

In the top panel of Figure \ref{fig:mpg} we show the membership probability calculated with the standard approach, without the parallax term, while on the bottom panel we show the results obtained including the parallax. In the left panels ($0\,\%<P<95\,\%$) we can see that there are considerably less sources with magnitudes in the range $10<G<19$ within $25\,\%<P<75\,\%$ if we account for the parallax distribution (372 sources in the top panel, 282 in the bottom), confirming that the discrimination between member and field stars is better.

\begin{figure}
    \centering
    \includegraphics[width=\columnwidth]{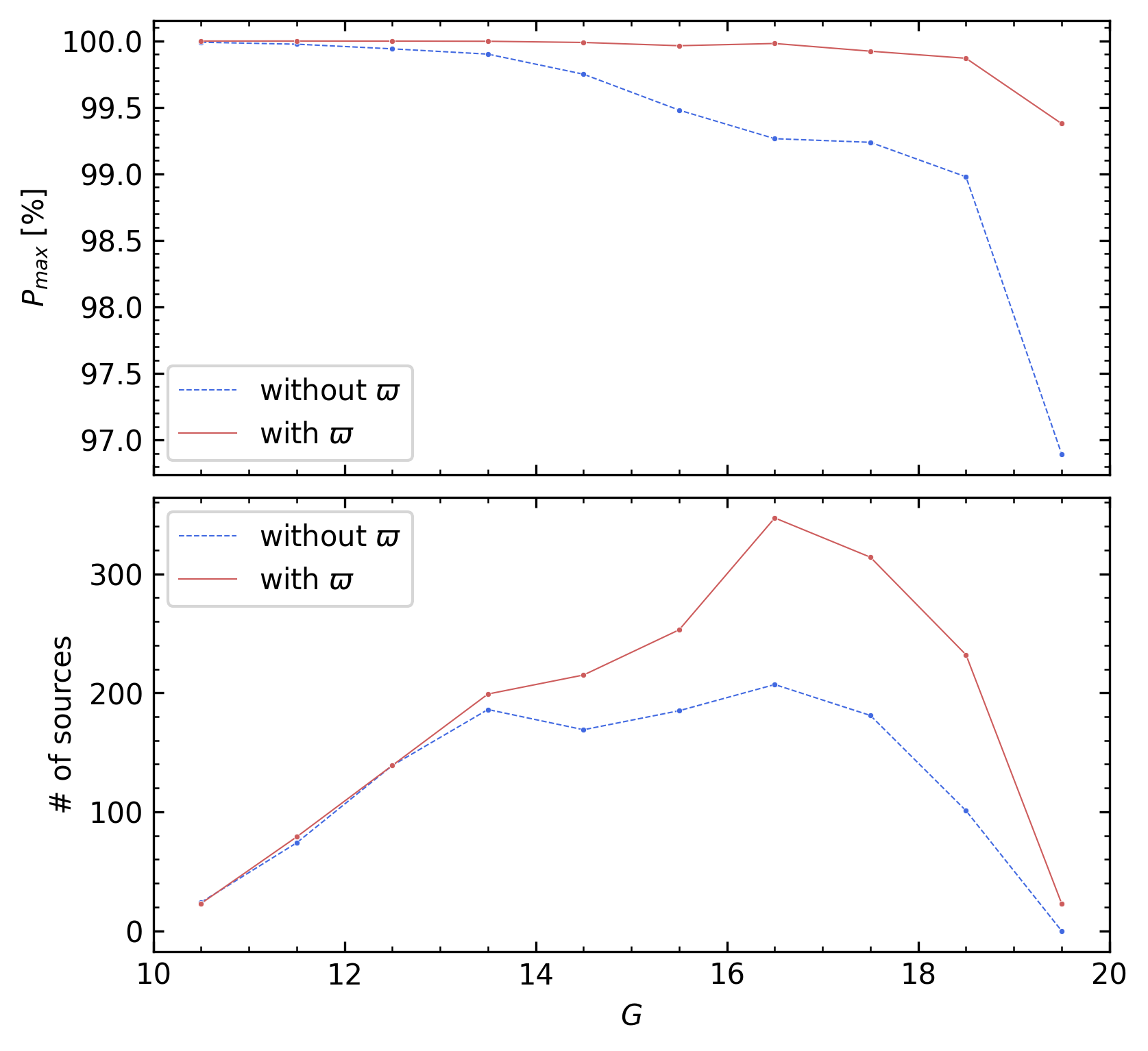}
    \caption{Top: maximum membership probability per magnitude bin. The dashed blue (solid red) line is the membership probability computed without (with) taking into account the parallax distribution. Bottom: number of sources per magnitude bin. The dashed blue (solid red) line is the number of sources with $P>97.5$\,\% ($P_\varpi>97.5$\,\% ).}
    \label{fig:p1p2}
\end{figure}

In Figure \ref{fig:p1p2} (top) we plotted the maximum membership probability per magnitude bin versus the Gaia $G$ magnitude. The blue and red lines represent the probability calculated with and without taking into account the parallax respectively.
This plot shows us that very high membership probabilities extend deeper when considering the parallax contribution in the calculation (about $1$\,mag at $P_{\rm{max}}=99$\,\%).

We then divided the sources in magnitude bins of $1$\,mag, and calculated the number of sources in each bin with $P>97.5$\,\%. In Figure \ref{fig:p1p2} (bottom) we plot the results: in blue we show the points obtained without accounting for the parallax, in red those obtained including the parallax in the membership calculation. It is clear that for $G \gtrsim 14$ we find more member stars if we use the parallax term. In total we found 1266 sources with $P\geq97.5$\,\%, and 1824 sources with $P_\varpi\geq97.5$\,\% in the region $10<G<20$, where $P_\varpi$ is the membership probability calculated with the formalism introduced in this work.

\begin{figure*}
    \centering
    \includegraphics[width=.9\textwidth]{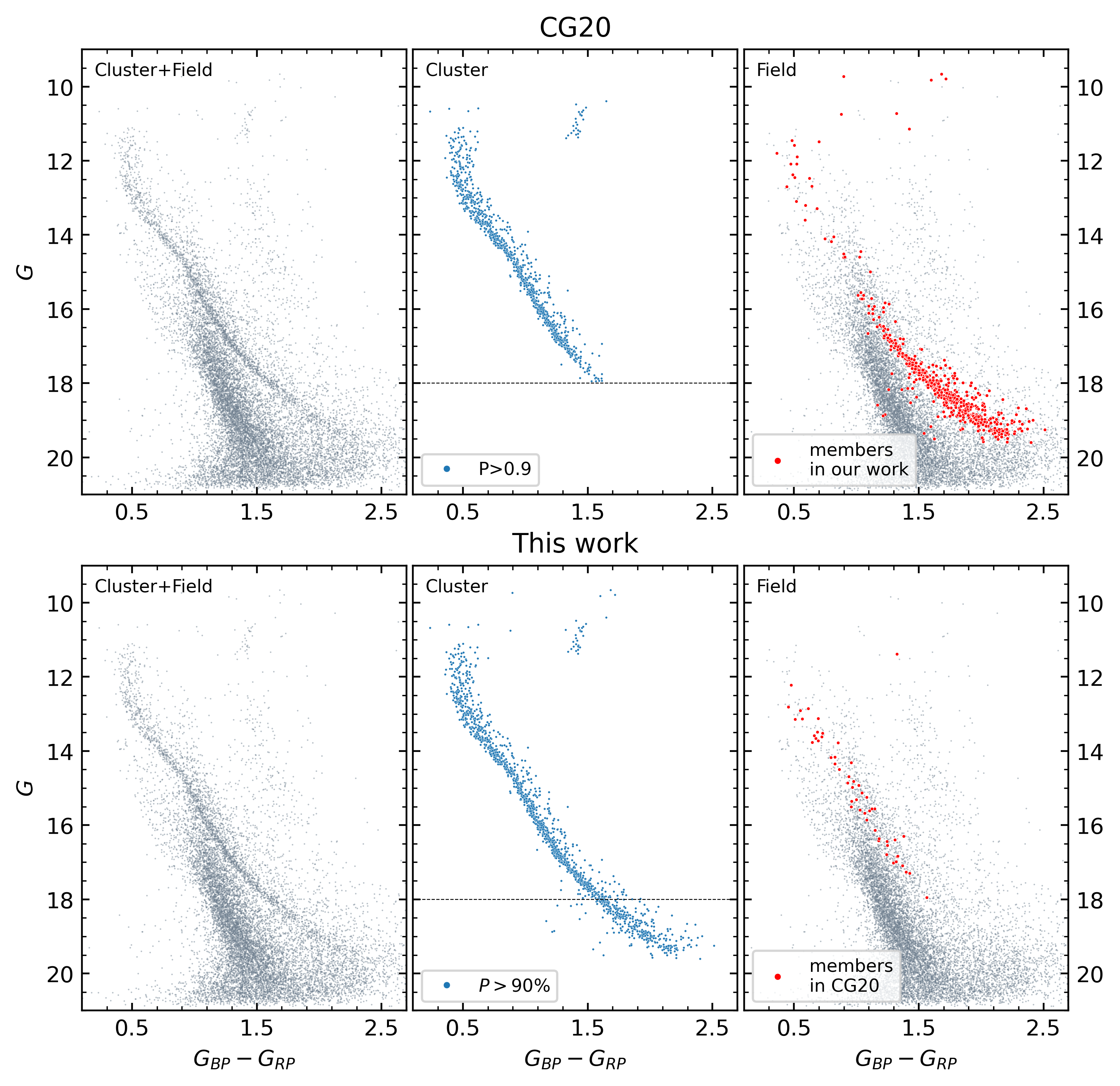}
    \caption{Color-magnitude diagram of the sources in M\,37 field of view: top panels show members identified by \citetalias{2020A&A...633A..99C}, while bottom panels show the members selected in this work. Left: all the sources. Center: sources with $P\geq0.9$\,\% (top) and $P_\varpi\geq90$\,\% (bottom). Right: sources with $P<0.9$\,\% (top) and $P_\varpi<90$\,\% (bottom); in the top (bottom) panel we highlighted in red the sources that passed the membership cut in the bottom (top) row. See text for more details.}
    \label{fig:cmd}
\end{figure*}

Figure \ref{fig:cmd} shows a comparison between \citetalias{2020A&A...633A..99C} and this work.
In the left column we show all the sources in the catalog; in this figure we limit the sample only to sources within a radius of 0.3\,deg 
\citep[slightly more than the cluster radius given by][]{2002A&A...389..871D} from the center of the cluster, as the number of field objects beyond this limit would overwhelm the plot, making the comparisons less clear.

In the central column we show the stars with $P\geq0.9$\, (top panel) and with $P_\varpi\geq90$\,\% (bottom panel). 
The clustering score given by  \citetalias{2020A&A...633A..99C} is provided only for stars brighter than $G=18$ (black dashed line) and according to the authors is a proxy for cluster membership probability. However the Main Sequence of M\,37 clearly extends (and it is well populated) also to fainter magnitudes than that limit.  
In the common region analysed by both works ($G\leq18$) we found about 200 extra sources with membership probability greater than 90\%, with respect to those with clustering score greater than 0.9 by  \citetalias{2020A&A...633A..99C}.
Nevertheless, the most interesting plots are shown in 
the right column, where we plotted the sources that did not pass the membership selection; in the top panel we highlighted in red the sources that did not pass the $P>0.9$ selection in the middle-top panel, but that passed the $P>90\%$ membership probability of the present work. 
Conversely, the stars in red in the bottom-right panel are those members according to $P>0.9$ in GC20, 
but not to the here derived $P>90\%$. 
Apart from the obvious improvement of the present work in finding members beyond $G>18$, we note a significant improvement in identifying members also in the magnitude range $16<G<18$. \\~

\section{Astrometric Parameters of M\,37}
\label{sec4}
\begin{figure}
    \centering
    \includegraphics[width=\columnwidth]{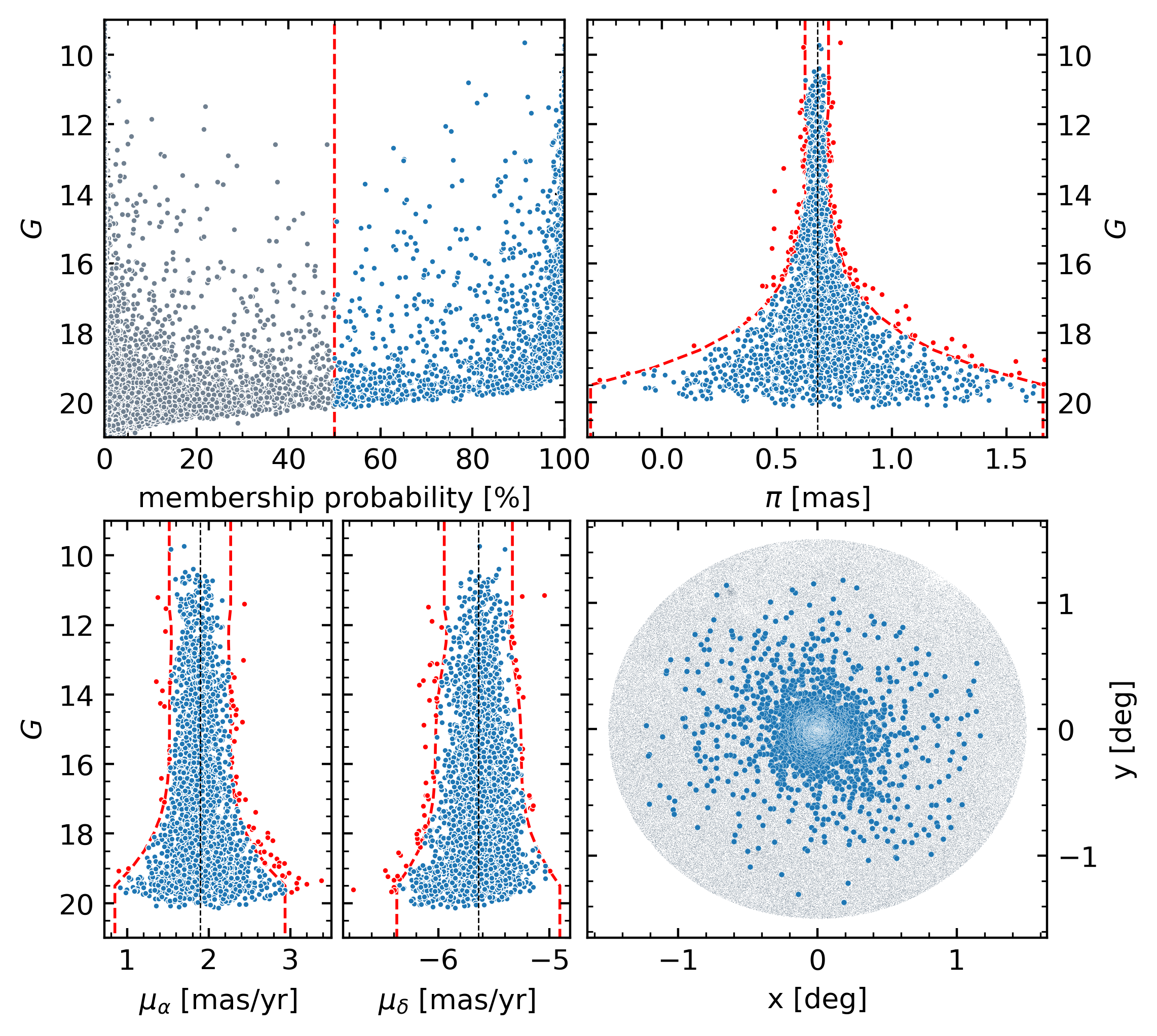}
    \caption{Members selection. Top left: membership probability for all the sources. We reject the stars with $P<50$\,\%. Top right: $G$ magnitude vs parallax. Here we reject the stars that fall outside the region delimited by the two red lines. The black dashed line represents the median parallax. Bottom left: proper motion of the sources vs their $G$ magnitude. We kept the sources between the red dashed lines. The vertical black line is the median proper motion. Bottom right: spatial distribution of the sources. Blue markers are the selected members of M\,37.}
    \label{fig:memb}
\end{figure}

To derive the mean astrometric parameters of M\,37 from the EDR3 catalog, we first need to select the most probable cluster members.
The selection procedure is illustrated in Figure \ref{fig:memb}. 
In the top-left panel we show the membership probability plotted against the $G$ band magnitude. We started by rejecting all the sources with membership probability lower than 50\%. 
Among these sources we rejected those falling outside the area delimited by the two red dashed line on the top-right panel. To define these red lines we proceeded as follows: first we divided the stars into $G$-magnitude bins of 0.5, for each bin we calculated the $3\sigma$-clipped median of the errors on the parallax given by the EDR3, and we took this median --multiplied by a factor of 2.5-- as the maximum error for members at the given $G$-magnitude.
We then define the red lines as a spline through these maxima. 
In the bottom-left panels we applied a similar cut, but we did not use the measurements errors on the proper motion from the Gaia catalog as the errors are much smaller than the intrinsic proper motion dispersion of cluster members (especially at the brighter magnitudes). 
Therefore, to define the widths of each bin we used instead the $68.27^{\rm th}$ of the observed residuals from the median (defined after a $3\sigma$-clipping), and again multiplied by a factor of 2.5. 
On the bottom-right panel we show the spatial distribution of the stars 
that passed all these four selections and which we then we consider as most probable members of M\,37.\\ 

We then further restrict this sample to the very best stars, requiring:
\begin{enumerate}
    \item $P_\varpi>99.5$\,\%, i.e., high confidence members; 
    \item magnitudes in all the three Gaia filters (no color trends); 
    \item $13\leq G\leq 15.4$, were the astrometric calibration 
    of the EDR3 catalog provide homogeneous errors \citep[cfr.][]{2021A&A...649A...5F};
    \item $\sigma_\varpi / \varpi < 0.1$, $\sigma_{\mu_{\alpha}} / \mu_\alpha < 0.1$ and $\sigma_{\mu_{\delta}} / \mu_\delta < 0.1$; 
    \item passing a number of quality cuts on the diagnostic parameters provided within the Gaia\,EDR3, 
    as done by \cite{2021ApJ...908L...5S}.
\end{enumerate}
Specifically, these applied quality-parameters cuts are:
\begin{description}
    \item \texttt{astrometric\_excess\_ noise}\,$<1$;
    \item \texttt{astrometric\_excess\_noise\_sig}\,$<=10$;
    \item \texttt{phot\_bp\_rp\_excess\_factor}\,$<1.6$;
    \item \texttt{phot\_proc\_mode}\,$=0$;
    \item \texttt{astrometric\_gof\_al}\,$<4$. 
\end{description}

\begin{figure}
    \centering
    \includegraphics[width=\columnwidth]{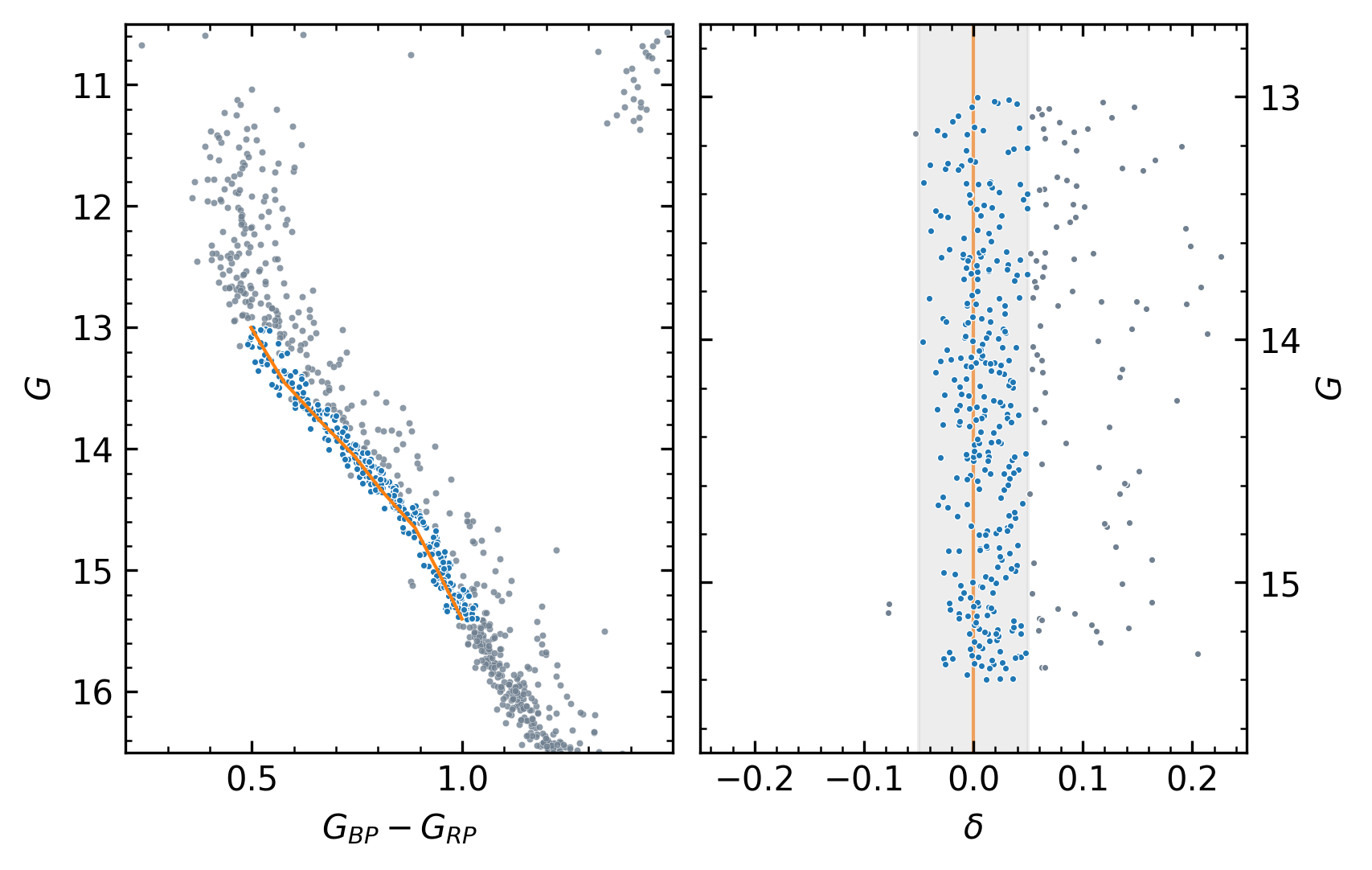}
    \caption{Left: color-magnitude diagram showing in blue the sources that we used to estimate the mean parameters of M\,37. Right: colour residuals from the fiducial for the stars in the selected sample; we used the sources in the shaded gray area which corresponds to $1\sigma$.}
    \label{fig:par_sel}
\end{figure}

After these selections we considered the color-magnitude diagram (CMD) of member stars (Figure \ref{fig:par_sel}, left panel) and applied a constraint in the $G$ vs $(G_{BP}-G_{RP})$ plane to exclude the region of the CMD populated by high-mass ratio photometric binaries (real or blends) that may have lower precision astrometry. This is achieved as follows: 
we divided the sample into $G$-magnitude bins of 0.3\,mag and we arbitrarily defined a specific colour for each bin as the 
$30^{\rm th}$ percentile of the colour distribution of the stars in the bin. 
We then interpolated these points at any given $G$-mag with a spline. 
The fiducial line defined in this way follows the bluer envelope of the Main-Sequence, as shown in the left panel of Figure \ref{fig:par_sel}. We then calculated the colour residuals $\delta$ from the fiducial and discarded the sources with |$\delta$|\,$>1\sigma$ (Figure \ref{fig:par_sel}, right panel).

With this tight selection of the very best measured and most likely members just defined for M\,37, 
we now proceed with our own derivation of the cluster mean astrometric parameters. 
We first compute the $3\sigma$-clipped median of $\varpi$, $\mu_\alpha$ and $\mu_\delta$ for each $G$-bin of 0.5\,mag, with $\sigma$ defined as the $68.27^{\rm th}$ percentile of the residuals around the median. The error associated with each bin is defined as $\epsilon_{k}=\sigma/\sqrt(N-1)$, with $N$ the number of sources in the bin. The values for the mean parallax and proper motions are calculated as a weighted mean through all the bins, with $1/\epsilon_k^2$ as weight.

\begin{figure}
    \centering
    \includegraphics[width=\columnwidth]{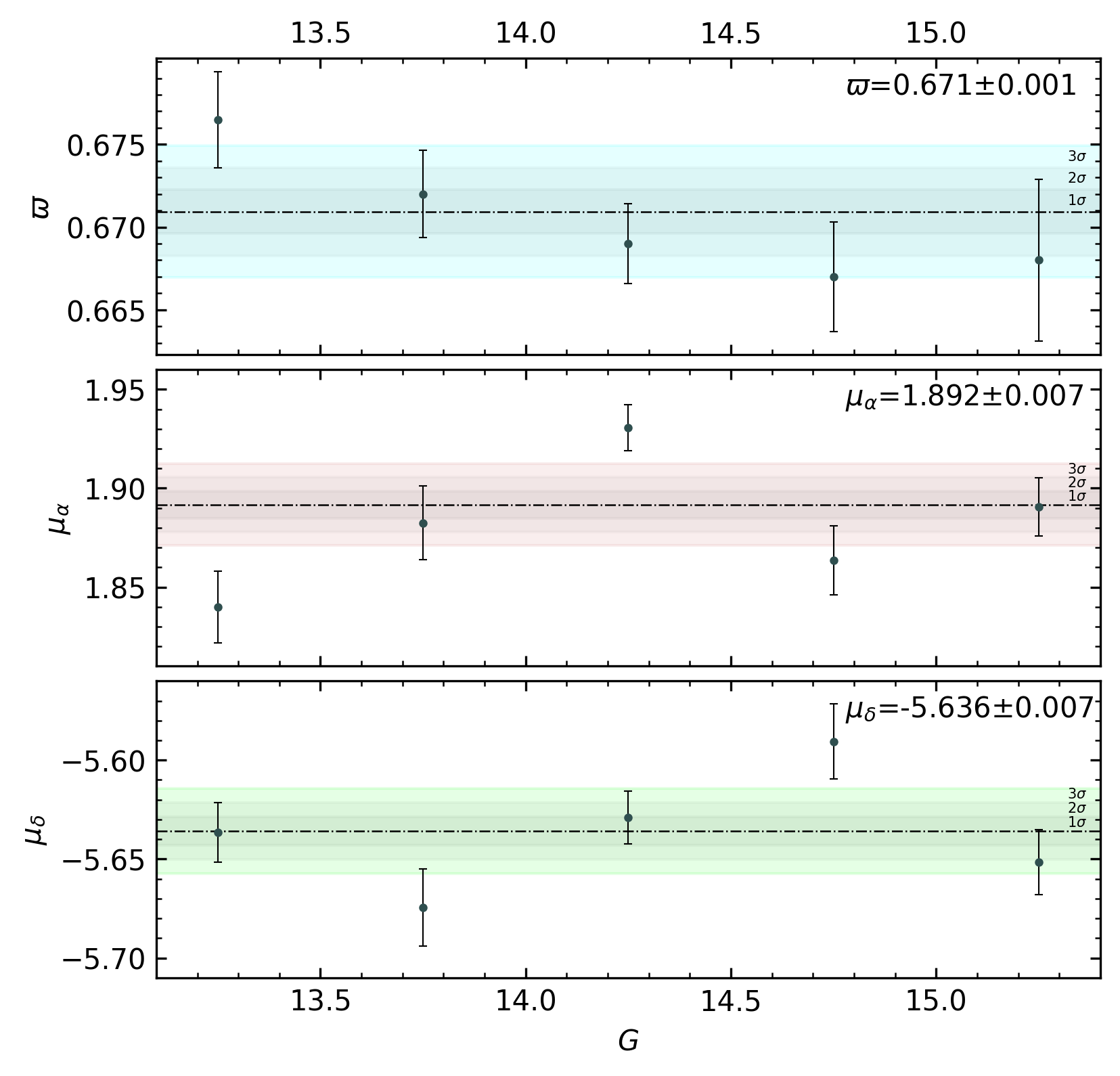}
    \caption{Mean values of the parallax (top) and proper motion components (middle and bottom) in each magnitude bin; the dash-dotted line is the overall weighted mean value, which is reported on the top right corner of each panel and in Table\,\ref{tab:m37par}.}
    \label{fig:est}
\end{figure}

~\\~ 

As the astrometric parameters for M\,37 are now better determined, thanks of the use of 
EDR3 (instead of being based on DR2 as in CG20) and the improved memberships,  
it makes more sense to use the newly determined cluster's mean parameters as starting values for our algorithm and to re-determine the membership probabilities.
Therefore, we repeated the analysis just discussed to derive our final estimate of the mean proper motion and parallax of M\,37. The values of mean parallax and proper motions for each magnitude bin are plotted in Fig.\,\ref{fig:est}, with the weighted mean through all the bins shown on the top right of each panel. These final values are also reported in Table\,\ref{tab:m37par}. We point out that neglecting the last selection on the CMD (displayed in Figure \ref{fig:par_sel}) our estimates do not change significantly (less than $0.3\sigma$).
 
Finally, \cite{2021A&A...649A...4L} found that EDR3 parallaxes of sources identified as quasars are systematically offset from the expected distribution around zero by a few tens of microarcseconds. 
They give an attempt to account for this offset which depends non trivially on the magnitude, colour, and ecliptic latitude of the source. We used their \texttt{Python} code to correct the parallaxes of M\,37 members and then recomputed the mean value. However, as they point out in their work, this correction is still under development and has problems, which seems to be supported by the disagreement with the expected values of $\sim 20\, \mu$as \citep[cfr. Figure 5 of][]{2021A&A...649A...4L}. We report also in Table\,\ref{tab:m37par} this bias-corrected value, as ${\varpi}^{\rm L21}$.

As a final note, while the absolute value of the parallax has not a direct effect on the membership probabilities, which is mostly a differential computation, it would still be good to have an indication of the systematic error in the just derived parallax. 
Therefore, we can conservatively associate a maximal error of $|\varpi-{\varpi}^{\rm L21}| = 0.043$\,mas, to the absolute parallax of M\,37 derived from Gaia\,EDR3: 
$0.671\pm0.001\pm0.043$\,mas, i.e., corresponding to a distance of 1.5$\pm$0.1\,kpc. 

\begin{table}
\centering
\caption{M\,37 mean parallax and proper motion. The value of $\varpi^{\rm L21}$ 
is the parallax corrected for the bias as is \protect\cite{2021A&A...649A...4L}.}
    \begin{tabular}{lcc}
    \hline
    \hline
    Parameter & value & unit \\
    \hline 
         $\varpi$            & $ 0.671 \pm 0.001$ & mas \\ 
         $\varpi^{\rm L21}$  & $ 0.629 \pm 0.001$ & mas \\ 
         $\mu_\alpha$        & $ 1.892 \pm 0.007$ & mas\,yr${-1}$\\ 
         $\mu_\delta$        & $-5.636 \pm 0.007$ & mas\,yr${-1}$\\ 
    \hline
    \end{tabular}
\label{tab:m37par}
\end{table}
%

\section{Catalog of M\,37}
\label{sec5}
As part of this work, we electronically release as Supporting Information on the Journal a catalog containing the Gaia\,EDR3 \texttt{source\,ID} and $P_\varpi$ (the membership probability calculated with the formalism presented in this work).

\section{Summary}
\label{sec6}
%
In this paper we presented a simple term, which involve parallaxes, to extend the classical method for computing cluster-membership probabilities based \textit{only} on 
proper motions and spatial distributions.  
The proposed new formalism, therefore, takes into account the full-astrometric information to compute memberships. 
Although currently this method suite only data provided by the Gaia\,EDR3 catalog, in principle this formalism could be adopted also to future other 5-parameters high-precision astrometric catalogs, or possibly to extensions of the Gaia astrometry to fainter magnitudes 
exploiting superior instruments capabilities \citep[e.g. using \textit{Hubble Space Telescope} observations as in][]{2018MNRAS.481.5339B,2020MNRAS.494.2068B}. 
We also note that employing relative instead of absolute parallaxes would not affect the membership probabilities as parallaxes enter only as a relative quantity in the calculations, nor would make any difference to add corrections for the systematic errors, such as those described in \cite{2021A&A...649A...4L} for Gaia\,EDR3.

We successfully applied this formalism to the case of the close-by open cluster M\,37, and release the derived membership probabilities. Results show that the new term allow us to better separate cluster members from field stars at all the magnitudes. We finally used the here-derived list of members to give a new estimate of the astrometric parameters of the cluster.

Future improvements of the method might combine the photometric information. Indeed, especially on wide open clusters with sparse densities, or in their outskirts in general, field objects might survive even tights membership probability selections, incidentally having same distance and motion of the clusters. 
Other future works might also include a term that would take into account the velocity along the line-of-sight of the sources (commonly referred to as \textit{radial velocities} in spectroscopy), when available. As Gaia radial velocities have not a great precision (200-300\,m\,s$^{-1}$ at best, up to 2.5\,km\,s$^{-1}$), nor they extend to sufficiently faint magnitudes ($G$ in the range 4-13), we ignored this term in this paper, which is focused on the astrometric parameters only.

\section*{Acknowledgements}

This work has made use of data from the European Space Agency (ESA) mission
{\it Gaia} (\url{https://www.cosmos.esa.int/gaia}), processed by the {\it Gaia}
Data Processing and Analysis Consortium (DPAC,
\url{https://www.cosmos.esa.int/web/gaia/dpac/consortium}). Funding for the DPAC
has been provided by national institutions, in particular the institutions
participating in the {\it Gaia} Multilateral Agreement.
The authors acknowledge support by MIUR under PRIN program \#2017Z2HSMF and 
by PRIN-INAF\,2019 under program \#10-Bedin.
%

\section*{Data Availability}

This work has made use of data from the European Space Agency (ESA) mission
{\it Gaia} (\url{https://www.cosmos.esa.int/gaia}), processed by the {\it Gaia}
Data Processing and Analysis Consortium (DPAC,
\url{https://www.cosmos.esa.int/web/gaia/dpac/consortium}).



\bibliographystyle{mnras}
\bibliography{bibliography} 

\begin{thebibliography}{}
\makeatletter
\relax
\def\mn@urlcharsother{\let\do\@makeother \do\$\do\&\do\#\do\^\do\_\do\%\do\~}
\def\mn@doi{\begingroup\mn@urlcharsother \@ifnextchar [ {\mn@doi@}
  {\mn@doi@[]}}
\def\mn@doi@[#1]#2{\def\@tempa{#1}\ifx\@tempa\@empty \href
  {http://dx.doi.org/#2} {doi:#2}\else \href {http://dx.doi.org/#2} {#1}\fi
  \endgroup}
\def\mn@eprint#1#2{\mn@eprint@#1:#2::\@nil}
\def\mn@eprint@arXiv#1{\href {http://arxiv.org/abs/#1} {{\tt arXiv:#1}}}
\def\mn@eprint@dblp#1{\href {http://dblp.uni-trier.de/rec/bibtex/#1.xml}
  {dblp:#1}}
\def\mn@eprint@#1:#2:#3:#4\@nil{\def\@tempa {#1}\def\@tempb {#2}\def\@tempc
  {#3}\ifx \@tempc \@empty \let \@tempc \@tempb \let \@tempb \@tempa \fi \ifx
  \@tempb \@empty \def\@tempb {arXiv}\fi \@ifundefined
  {mn@eprint@\@tempb}{\@tempb:\@tempc}{\expandafter \expandafter \csname
  mn@eprint@\@tempb\endcsname \expandafter{\@tempc}}}

\bibitem[\protect\citeauthoryear{{Balaguer-N{\'u}nez}, {Tian}  \&
  {Zhao}}{{Balaguer-N{\'u}nez} et~al.}{1998}]{1998A&AS..133..387B}
{Balaguer-N{\'u}nez} L.,  {Tian} K.~P.,   {Zhao} J.~L.,  1998, \mn@doi [\aaps]
  {10.1051/aas:1998324}, \href
  {https://ui.adsabs.harvard.edu/abs/1998A&AS..133..387B} {133, 387}

\bibitem[\protect\citeauthoryear{{Baumgardt}, {Dettbarn}  \&
  {Wielen}}{{Baumgardt} et~al.}{2000}]{2000A&AS..146..251B}
{Baumgardt} H.,  {Dettbarn} C.,   {Wielen} R.,  2000, \mn@doi [\aaps]
  {10.1051/aas:2000362}, \href
  {https://ui.adsabs.harvard.edu/abs/2000A&AS..146..251B} {146, 251}

\bibitem[\protect\citeauthoryear{{Bedin} \& {Fontanive}}{{Bedin} \&
  {Fontanive}}{2018}]{2018MNRAS.481.5339B}
{Bedin} L.~R.,  {Fontanive} C.,  2018, \mn@doi [\mnras]
  {10.1093/mnras/sty2626}, \href
  {https://ui.adsabs.harvard.edu/abs/2018MNRAS.481.5339B} {481, 5339}

\bibitem[\protect\citeauthoryear{{Bedin} \& {Fontanive}}{{Bedin} \&
  {Fontanive}}{2020}]{2020MNRAS.494.2068B}
{Bedin} L.~R.,  {Fontanive} C.,  2020, \mn@doi [\mnras]
  {10.1093/mnras/staa540}, \href
  {https://ui.adsabs.harvard.edu/abs/2020MNRAS.494.2068B} {494, 2068}

\bibitem[\protect\citeauthoryear{{Bellini} et~al.,}{{Bellini}
  et~al.}{2009}]{2009A&A...493..959B}
{Bellini} A.,  et~al., 2009, \mn@doi [\aap] {10.1051/0004-6361:200810880},
  \href {https://ui.adsabs.harvard.edu/abs/2009A&A...493..959B} {493, 959}

\bibitem[\protect\citeauthoryear{{Cantat-Gaudin} \& {Anders}}{{Cantat-Gaudin}
  \& {Anders}}{2020}]{2020A&A...633A..99C}
{Cantat-Gaudin} T.,  {Anders} F.,  2020, \mn@doi [\aap]
  {10.1051/0004-6361/201936691}, \href
  {https://ui.adsabs.harvard.edu/abs/2020A&A...633A..99C} {633, A99}

\bibitem[\protect\citeauthoryear{{Cantat-Gaudin} et~al.,}{{Cantat-Gaudin}
  et~al.}{2018}]{2018A&A...615A..49C}
{Cantat-Gaudin} T.,  et~al., 2018, \mn@doi [\aap]
  {10.1051/0004-6361/201731251}, \href
  {https://ui.adsabs.harvard.edu/abs/2018A&A...615A..49C} {615, A49}

\bibitem[\protect\citeauthoryear{{Castro-Ginard}, {Jordi}, {Luri}, {Julbe},
  {Morvan}, {Balaguer-N{\'u}{\~n}ez}  \& {Cantat-Gaudin}}{{Castro-Ginard}
  et~al.}{2018}]{2018A&A...618A..59C}
{Castro-Ginard} A.,  {Jordi} C.,  {Luri} X.,  {Julbe} F.,  {Morvan} M.,
  {Balaguer-N{\'u}{\~n}ez} L.,   {Cantat-Gaudin} T.,  2018, \mn@doi [\aap]
  {10.1051/0004-6361/201833390}, \href
  {https://ui.adsabs.harvard.edu/abs/2018A&A...618A..59C} {618, A59}

\bibitem[\protect\citeauthoryear{{Dias}, {Alessi}, {Moitinho}  \&
  {L{\'e}pine}}{{Dias} et~al.}{2002}]{2002A&A...389..871D}
{Dias} W.~S.,  {Alessi} B.~S.,  {Moitinho} A.,   {L{\'e}pine} J.~R.~D.,  2002,
  \mn@doi [\aap] {10.1051/0004-6361:20020668}, \href
  {https://ui.adsabs.harvard.edu/abs/2002A&A...389..871D} {389, 871}

\bibitem[\protect\citeauthoryear{{Fabricius} et~al.,}{{Fabricius}
  et~al.}{2021}]{2021A&A...649A...5F}
{Fabricius} C.,  et~al., 2021, \mn@doi [\aap] {10.1051/0004-6361/202039834},
  \href {https://ui.adsabs.harvard.edu/abs/2021A&A...649A...5F} {649, A5}

\bibitem[\protect\citeauthoryear{{Gagn{\'e}} et~al.,}{{Gagn{\'e}}
  et~al.}{2018}]{2018ApJ...856...23G}
{Gagn{\'e}} J.,  et~al., 2018, \mn@doi [\apj] {10.3847/1538-4357/aaae09}, \href
  {https://ui.adsabs.harvard.edu/abs/2018ApJ...856...23G} {856, 23}

\bibitem[\protect\citeauthoryear{{Gaia Collaboration} et~al.,}{{Gaia
  Collaboration} et~al.}{2016}]{2016A&A...595A...1G}
{Gaia Collaboration} et~al., 2016, \mn@doi [\aap]
  {10.1051/0004-6361/201629272}, \href
  {https://ui.adsabs.harvard.edu/abs/2016A&A...595A...1G} {595, A1}

\bibitem[\protect\citeauthoryear{{Gaia Collaboration} et~al.,}{{Gaia
  Collaboration} et~al.}{2021}]{2021A&A...649A...1G}
{Gaia Collaboration} et~al., 2021, \mn@doi [\aap]
  {10.1051/0004-6361/202039657}, \href
  {https://ui.adsabs.harvard.edu/abs/2021A&A...649A...1G} {649, A1}

\bibitem[\protect\citeauthoryear{{Kozhurina-Platais}, {Girard}, {Platais}, {van
  Altena}, {Ianna}  \& {Cannon}}{{Kozhurina-Platais}
  et~al.}{1995}]{1995AJ....109..672K}
{Kozhurina-Platais} V.,  {Girard} T.~M.,  {Platais} I.,  {van Altena} W.~F.,
  {Ianna} P.~A.,   {Cannon} R.~D.,  1995, \mn@doi [\aj] {10.1086/117310}, \href
  {https://ui.adsabs.harvard.edu/abs/1995AJ....109..672K} {109, 672}

\bibitem[\protect\citeauthoryear{{Lindegren} et~al.,}{{Lindegren}
  et~al.}{2021a}]{2021A&A...649A...2L}
{Lindegren} L.,  et~al., 2021a, \mn@doi [\aap] {10.1051/0004-6361/202039709},
  \href {https://ui.adsabs.harvard.edu/abs/2021A&A...649A...2L} {649, A2}

\bibitem[\protect\citeauthoryear{{Lindegren} et~al.,}{{Lindegren}
  et~al.}{2021b}]{2021A&A...649A...4L}
{Lindegren} L.,  et~al., 2021b, \mn@doi [\aap] {10.1051/0004-6361/202039653},
  \href {https://ui.adsabs.harvard.edu/abs/2021A&A...649A...4L} {649, A4}

\bibitem[\protect\citeauthoryear{{Monteiro}, {Dias}, {Moitinho},
  {Cantat-Gaudin}, {L{\'e}pine}, {Carraro}  \& {Paunzen}}{{Monteiro}
  et~al.}{2020}]{2020MNRAS.499.1874M}
{Monteiro} H.,  {Dias} W.~S.,  {Moitinho} A.,  {Cantat-Gaudin} T.,
  {L{\'e}pine} J.~R.~D.,  {Carraro} G.,   {Paunzen} E.,  2020, \mn@doi [\mnras]
  {10.1093/mnras/staa2983}, \href
  {https://ui.adsabs.harvard.edu/abs/2020MNRAS.499.1874M} {499, 1874}

\bibitem[\protect\citeauthoryear{{Nardiello} et~al.,}{{Nardiello}
  et~al.}{2018}]{2018MNRAS.481.3382N}
{Nardiello} D.,  et~al., 2018, \mn@doi [\mnras] {10.1093/mnras/sty2515}, \href
  {https://ui.adsabs.harvard.edu/abs/2018MNRAS.481.3382N} {481, 3382}

\bibitem[\protect\citeauthoryear{{Robichon}, {Arenou}, {Mermilliod}  \&
  {Turon}}{{Robichon} et~al.}{1999}]{1999A&A...345..471R}
{Robichon} N.,  {Arenou} F.,  {Mermilliod} J.~C.,   {Turon} C.,  1999, \aap,
  \href {https://ui.adsabs.harvard.edu/abs/1999A&A...345..471R} {345, 471}

\bibitem[\protect\citeauthoryear{{Sanders}}{{Sanders}}{1971}]{1971A&A....14..226S}
{Sanders} W.~L.,  1971, \aap, \href
  {https://ui.adsabs.harvard.edu/abs/1971A&A....14..226S} {14, 226}

\bibitem[\protect\citeauthoryear{{Scalco} et~al.,}{{Scalco}
  et~al.}{2021}]{2021MNRAS.505.3549S}
{Scalco} M.,  et~al., 2021, \mn@doi [\mnras] {10.1093/mnras/stab1476}, \href
  {https://ui.adsabs.harvard.edu/abs/2021MNRAS.505.3549S} {505, 3549}

\bibitem[\protect\citeauthoryear{{Soltis}, {Casertano}  \& {Riess}}{{Soltis}
  et~al.}{2021}]{2021ApJ...908L...5S}
{Soltis} J.,  {Casertano} S.,   {Riess} A.~G.,  2021, \mn@doi [\apjl]
  {10.3847/2041-8213/abdbad}, \href
  {https://ui.adsabs.harvard.edu/abs/2021ApJ...908L...5S} {908, L5}

\bibitem[\protect\citeauthoryear{{Tian}, {Zhao}, {Shao}  \& {Stetson}}{{Tian}
  et~al.}{1998}]{1998ycat..41310089T}
{Tian} K.~P.,  {Zhao} J.~L.,  {Shao} Z.~Y.,   {Stetson} P.~B.,  1998, VizieR
  Online Data Catalog, \href
  {https://ui.adsabs.harvard.edu/abs/1998yCat..41310089T} {pp J/A+AS/131/89}

\bibitem[\protect\citeauthoryear{{Vasilevskis}, {Klemola}  \&
  {Preston}}{{Vasilevskis} et~al.}{1958}]{1958AJ.....63..387V}
{Vasilevskis} S.,  {Klemola} A.,   {Preston} G.,  1958, \mn@doi [\aj]
  {10.1086/107787}, \href
  {https://ui.adsabs.harvard.edu/abs/1958AJ.....63..387V} {63, 387}

\bibitem[\protect\citeauthoryear{{Yadav} et~al.,}{{Yadav}
  et~al.}{2008}]{2008A&A...484..609Y}
{Yadav} R.~K.~S.,  et~al., 2008, \mn@doi [\aap] {10.1051/0004-6361:20079245},
  \href {https://ui.adsabs.harvard.edu/abs/2008A&A...484..609Y} {484, 609}

\makeatother
\end{thebibliography}

\bsp	
\label{lastpage}
\end{document}